# Spin coherence time analytical estimations

Yuri Orlov (Cornell University)

5 June 2015

## Abstract

Section I presents a variety of analytical estimations related to spin coherence time (SCT) in a purely electric frozen-spin ring. The main result is that, in the case of $m > 0$ and vertical oscillations only, the kinetic energy equilibrium shift equals zero, that is, SCT does not depend on these oscillations. Section II contains additional information on this case concerning terminology, electric field definition and vertical oscillations.

## Introduction

This paper combines my talk at the 1-5 October 2012 workshop "EDM Searches at Storage Rings" ECT Trento, Italy [1] and my 23 January 2013 Storage Ring EDM Collaboration Note adding details to that talk [2]. Both referred to simulations done by S. Hacıömeroğlu and Y. K. Semertzidis in 2011 [3]. These have been updated, yielding more precise results. I reference those updated simulations here [4], and have now included all relevant plots in an appendix.

**Section I**

In the presence of synchrotron (and betatron) stability, spin coherence time is defined by a non-linear part of the particle deviations from the equilibrium. All degrees of freedom in this part are usually interconnected. What is unusual in an electric ring is that all degrees of freedom are strongly interconnected even in the linear approximation.

In a purely electric frozen-spin ring, the designed particle momentum is magic, $p/mc = 1/\sqrt{a}$. The designed g-2 frequency equals zero,

$$\vec{\omega}_a = \frac{e}{mc}\left[a - \left(\frac{mc}{p}\right)^2\right]\vec{\beta}\times\vec{E} = 0 . \qquad (1)$$

The central SCT problem is the magnitude of this frequency spread, $\Delta\omega_a$, caused by a shift and/or spread of the particles' momenta. The main linear effect of the initial $\Delta p$-spread is cancelled (on average in time) in the regime of synchrotron oscillations of $\Delta p$, so our concern is non-linear effects. For example, being in the denominator of $1/p^2$, the linear oscillations produce non-linear effects.

The Lorentz eq. for the energy component of the 4-momentum, $\gamma = dt/d\tau$, with $\tau$ for proper time,

$$d\gamma/d\tau = \frac{eE_R}{mc}dx/d\tau + \frac{eE_V}{mc}dy/d\tau + \frac{eE_L}{mc}ds/d\tau, \qquad (2)$$

connects the particle energy with all other degrees of freedom, $x$ (radial), $y$ (vertical) and $s$ (longitudinal), since all components of the electric fields are present in the Storage Ring

EDM Collaboration ring. In their simulations, Selçuk Hacıömeroğlu and Yannis Semertzidis use [3]:

$$E_R = \frac{E_0}{(1+x/R)^{1+m}} \left[ 1 - \frac{m(m+2)}{2} \frac{y^2}{R^2} + ... \right] ; \tag{3}$$

$$E_V = \frac{E_0}{(1+x/R)^{1+m}} \left[ m\frac{y}{R} - ... \right]. \tag{4}$$

In a frozen-spin EDM ring, we will use a very small $m$, so

$$E_R = \frac{E_0}{(1+x/R)} \left( 1 - m\frac{y^2}{R^2} \right), \quad E_V = \frac{E_0}{(1+x/R)} m\frac{y}{R} \tag{5}$$

is sufficient. I will not take into account the straight sections. In the linear approximation, Eq. (2) becomes

$$\frac{d\Delta\gamma/\gamma_0}{dt} = -\beta_0^2 \frac{dx/R}{dt} - \frac{eV\omega_C}{2\pi\gamma_0 mc^2}\phi, \text{ linear.} \tag{6}$$

$\phi$ is the synchrotron phase, $\omega_C$ the revolution frequency. In the linear approximation, we can consider $x/R$ and $\Delta\gamma/\gamma_0$ as sums of the fast (betatron) and slow (synchrotron) oscillations.

In the linear approximation of betatron oscillations, Eq. (6) with $\phi = 0$,

$$\gamma - \gamma_0 = -\gamma_0\beta_0^2(x/R). \tag{7}$$

This $\gamma - x$ interconnection, non-existent in magnetic rings, should be taken into account in the Lorenz eq. for $x$:

$$\frac{d^2x/R}{d\tau^2} - \frac{c^2}{R^2} \frac{(1+x/R)(ds/d\tau)^2}{R} = \frac{eE_R}{mc}\gamma, \tag{8}$$

where the second term describes the centrifugal force;

$$(ds/d\tau)^2 = \frac{\gamma^2 - 1 - (dx/d\tau)^2 - (dy/d\tau)^2}{(1+x/R)^2}. \tag{9}$$

In the linear approximation [5] for the case of $m = 0$,

$$\frac{d^2(x/R)}{dt^2} + \frac{c^2}{R^2}\beta_0^2(2-\beta_0^2)\left(\frac{x}{R}\right) = 0. \tag{10}$$

Here, $c^2\beta^2/R^2 = \omega_C^2$; $(2-\beta^2) = \nu_R^2$ is the radial focusing tune. In this case of the field index $m = 0$, the radial focusing is provided only by the momentum radial acceleration caused by the radial electric field.

In the linear approximation of synchrotron oscillations,

$$d\phi/dt = (h\omega_C)(x/R) - (h\omega_C/\gamma_0^2\beta_0^2)(\Delta\gamma/\gamma_0) = h\omega_C\frac{\alpha - 1/\gamma_0^2}{\beta_0^2}(\Delta\gamma/\gamma_0),$$

with $\alpha = 1$ [5].

Suppose we introduce some arbitrary initial conditions for $x/R$ and $\Delta\gamma/\gamma_0$. Generally, they will not satisfy the relation described by Eq. (6). As a result, only some of the particle oscillations will be betatron oscillations described by Eq. (11); the remainder inevitably belong to the synchrotron oscillations. This is why simulations expose a significant component of synchrotron oscillations (Appendix, Figs. 2, 3, 4, 5).

For the SCT investigation, we need to work in the quadratic approximation. It follows from (1) and from simulations (see [3] and Appendix) that, in general, $\Delta\omega_a$ depends on perturbations of both factors, $[a - (mc)^2/p^2]$ and $(\vec{\beta} \times \vec{E})$ of Eq. (1). For example, in the perturbation

$$\Delta[a - (mc/p)^2] = a[2\Delta p/p - 3(\Delta p/p)^2] \tag{11}$$

we need to know not only the shift, $\delta p / p$, of the momentum from the designed equilibrium. We also need to know the linear $\Delta p / p$ oscillations, which, being in the denominator, produce non-zero $(\Delta p / p)^2$ and, in addition, may interfere with the oscillations of $x/R$ present in another factor.

The shift, $\delta p / p$, is a quadratic effect, a constant not covered by Eq. (6). Information about this constant can be obtained from the eq. for the revolution period, $T = L / c\beta$, extended to include all non-linear, quadratic terms. The time-averaged revolution period is fixed by the RF frequency,

$$\bar{T} = 2\pi h / \omega_{RF}. \qquad (12)$$

Another eq. can be derived by averaging the Lorentz non-linear Eq. (8) for $x / R$.

I use the relativistic invariant and covariant Lorentz equations, which are based on the Lagrange formalism. I will use this method, below, to calculate the SCT for a case of vertical oscillations only, and $m <<< 1$.

Assume the presence of only vertical oscillations, and the fields as in Eq. (5). The energy depends only on $y$ and, in accordance with this, Eq. (2)—where only derivatives are involved—can contain two unknown constant shifts, $\delta\gamma$ and $\delta x$, from the designed values, $\gamma = \gamma_0$ and $x = 0$ :

$$\Delta\gamma = \gamma - \gamma_0 = \delta\gamma - \gamma_0\beta_0^2 \frac{\delta x}{R} - (\beta^2\gamma)_0 m \frac{y^2}{2R^2}. \qquad (13)$$

The non-linear equation of motion for radial coordinates can be written as:

$$\frac{d^2}{d\tau^2}\left(\frac{x}{R}\right) = \frac{(c/R)^2}{(1+\delta x/R)}\left\{(1-\vartheta_y^2)[\gamma_0^2 - 1 + 2\gamma_0(\gamma - \gamma_0)] - (\beta\gamma)_0^2[1 - m(y/R)^2][1 + \frac{(\gamma - \gamma_0)}{\gamma_0}]\right\} .$$

$$(14)$$

Now, taking into account that $(dy/d\tau)^2/(\beta\gamma)_0^2 = \vartheta_0^2 = m(y/R)^2$, the term proportional to $\vartheta_y^2$ cancels the term proportional to $(y/R)^2$. If we average this eq. over a long time, $\tau$, then the left side of (14) goes to zero. The obtained new equation is satisfied only if $(\gamma - \gamma_0) = 0$, so $\overline{\Delta p/p} = 0$ in Eq. (11). In the quadratic approximation, $\overline{\Delta p/p^2}$ also equals zero since $\Delta p$, $\Delta\gamma$ do not contain linearly oscillating terms. (By our assumption, the horizontal oscillations are absent.) Thus, the spin coherence is not perturbed by the particle vertical oscillations, SCT $\sim \infty$, at m<<1. This conclusion for vertical oscillations is consistent with a big SCT in simulation results (see Appendix, Fig. 1). And, according to Eq. (13), the equilibrium position and/or energy equilibrium are shifted.

*Remark 1*. If the quadratic terms in Eq. (14) did not cancel one another but were summarized instead, then $(\gamma - \gamma_0) \neq 0$:

$$-2v_z^2\left(\frac{z}{R}\right)^2(\gamma\beta)^2 = \frac{\Delta\gamma}{\gamma}\left[(\beta\gamma)_0^2 - 2\gamma_0^2\right] = -\gamma_0^2\beta_0^2\left(2-\beta^2\right)\frac{\Delta p}{p},$$

so

$$2\frac{\Delta p}{p} = \frac{4m}{\left(2-\beta^2\right)}\left(\frac{z}{R}\right)^2; \quad \text{with } m = 0.04,\ z = 2cm,\ R = 40m,\ T_c = 1.4\times10^{-6}s,\ a = 1.8,$$

$$\omega_a = \frac{e}{mc}E\beta\alpha \cdot \frac{4m}{2-\beta^2}\left(\frac{z}{R}\right)^2 = \omega_c\frac{4m}{\gamma\left(2-\beta^2\right)}\left(\frac{z}{R}\right)^2 \sim 0.1\ rad/s, \text{ vertical}.$$

In Appendix, Fig. 1, $1.6\times10^{-3} = 0.0016\,rad/s$, some 100 times smaller than this estimate of what would be if the quadratic terms in (14) did not cancel one another.

*Remark 2*. Radial oscillations, $\phi = 0$, $y = 0$, $m \to 0$ : a potential complication.

$$(\gamma - \gamma_0) = \delta\gamma - \gamma_0\beta_0^2\frac{\delta x}{R} - \gamma_0\beta_0^3\left[\frac{x}{R} - \frac{1}{2}\left(\frac{x}{R}\right)^2\right] + 0(m); \qquad (15)\ \text{OK}$$

$x$ here is a fast <u>oscillating mode</u>, $\delta x =$ constant.

$$(\gamma - \gamma_0)^2 = \left(\gamma_0\beta_0^2\right)^2\left(\frac{x}{R}\right)^2 + \text{ higher than second approx. terms.} \qquad (16)\ \text{OK}$$

$$\frac{d^2(x/R)}{d\tau^2} = \left(\frac{c}{R}\right)^2\frac{1}{(1+x/R)}\left\{\left(1-\theta_x^2\right)\left[(\gamma\beta)_0^2 + 2\gamma_0^2\left(\frac{\gamma-\gamma_0}{\gamma_0}\right) + (\gamma-\gamma_0)^2\right] - (\beta\gamma)_0^2\left(1+\frac{\gamma-\gamma_0}{\gamma_0}\right)\right\}$$

with the condition, $\bar{\omega}_l = \frac{c\beta_0}{R}$. Further, $\overline{\theta_x^2} = \left(2-\beta^2\right)\overline{(x/R)^2}$. $\qquad (17)\ \text{OK}$

What is $(x/R)(\Delta\gamma/\gamma)$ when $m \neq 0$? According to (15),

$$\left(\frac{x}{R}\right)\left(\frac{\Delta\gamma}{\gamma}\right) = -\gamma_0^2\beta_0^2\left(\frac{x}{R}\right)^2 + \text{ higher terms.}$$

Then, from the averaged (17),

$$\frac{\overline{\Delta\gamma}}{\gamma}\gamma_0^2\left(2-\beta_0^2\right) = -(\gamma\beta)_0^2\overline{(x/R)^2}. \qquad (18)$$

Eq. 18 IS NOT OK. The averaging is not well-defined. For example:

$$\overline{\frac{d^2x}{d\tau^2}} = \frac{1}{\tau}\int_0^\tau\frac{d^2x}{d\tau^2}d\tau = 0.$$

But, $\overline{\dfrac{d^2x}{d\tau^2}} = \dfrac{1}{t}\int_0^t dt\left(\dfrac{d^2x}{d\tau^2}\right) \neq 0$ : This averaging contains a "resonance" term proportional to

$\left(\dfrac{d\gamma}{dt}\right)\left(\dfrac{dx}{dt}\right) \sim \left(\dfrac{dx}{dt}\right)^2$ !

## Section II

*Terminology.* I use the four covariant equations for the Lorentz 4-force, $Du^i / D\tau = (e/mc)F^i_k u^k$, where $u^i = p^i / m$, $i = 0,1,2,3$, is the 4-velocity, but do not show these eqs. in this notation. Instead, I use the more familiar $\frac{dx}{dt}, \frac{dy}{dt}$, $\gamma$, etc. When working in the frame of these equations, I call $\gamma mc^2$ "the particle energy." Such terminology has been adopted by various authors. In [6], for example, $\varepsilon = m\gamma$ in Eq. (1.57), $d\varepsilon / dt = d / dt(m / \sqrt{1-v^2}) = ... = e\vec{E} \cdot \vec{v}$, is called "energy." This eq. for $\varepsilon$ is the same as my Eq. (2) for the fourth momentum component, $u^0 = dt / d\tau = \gamma$. Landau and Lifshitz [7] call $\gamma mc^2$ in the same equation "kinetic energy." Regardless of what it is called, Landau-Lifshitz's (as well as my) $d\gamma / dt$ includes the acceleration by all electric field components, radial, vertical and longitudinal, whether constant or changing in time, including the synchrotron RF field.

*Electric field.* The electric field described in Eqs. (3)-(4) is uniquely defined by the azimuthal symmetry and someone's free choice of the radial field as a function of $x$ in the plane $y=0$. In any choice, the value of the sextupole field component is crucially important for the duration of SCT [8]. In [1], I analyze the field chosen by Hacıömeroğlu and Semertzidis in [3]. My conclusion that the contribution of the vertical betatron oscillations to the spin decoherence equals zero (in the quadratic approximation) relates only to the electric field described in Eqs. (3) and (4).

*Vertical oscillations: the values of* $\delta\gamma$, $\delta x / R$. In the presence of RF, the equilibrium radius, equilibrium gamma (and hence velocity), and the magnitude of the electric field at the equilibrium are defined by three equations: in the case of the vertical oscillations only, they are defined by Eqs. (14), (13) and (12). Eq. (14) for the radial component of the (generalized) 4-momentum is written in the quadratic approximation for deviations from equilibrium, plus the assumption $m^2 << m <<< 1$ for the field index *m*. (This <<< notation has been borrowed from Gottfried and Yan's *Quantum Mechanics*.) This eq. is then averaged over a time much longer than the period of vertical oscillations [9]. It is taken into account that the shifts $\delta x$ (of the radius) and $\delta\gamma$ are quadratic or of a higher-order effect (in the presence of RF). It can be seen from Eq. (14) that the two vertical terms explicitly present in this eq. cancel each other. That is, the effect of the orbit lengthening by the vertical oscillations is cancelled by the sextupole component of the electric field. As a result,

$$\left\langle \vartheta_t^2 \right\rangle = m \left\langle y^2 / R^2 \right\rangle, \quad \left\langle \gamma \right\rangle = \gamma_0, \, \delta\gamma = 0. \tag{14'}$$

Thus, gamma is not shifted by vertical oscillations.

Solution (14') is not the end of the story. Eq. (13), which follows from the general Eq. (2), shows that in order to permit the solution $\gamma = \gamma_0$, the particle must shift its radius:

$$\left\langle \delta x / R \right\rangle = -\frac{m}{2} \left\langle y^2 / R^2 \right\rangle. \tag{13'}$$

Finally, let me show that Eqs. (13') and (14') are consistent with Eq. (12) in the presence of RF. Indeed, in the absence of radial and synchrotron oscillations, Eq. (12) is satisfied if the length of the orbit remains equal to the designed length, $L = L_0$, despite the vertical oscillations. From Eq. (3.6) of [8], $\Delta L = L - L_0 = 0$ is satisfied if

$$\langle x / R \rangle = -\langle \vartheta_y^2 / 2 \rangle, \tag{12'}$$

which is consistent with (13') and (14'). This consistency confirms my conclusion (page 6, above) about SCT in the presence of vertical oscillations.

## Acknowledgments

I would like to thank Selçuk Hacıömeroğlu and Yannis K. Semertzidis for their contributions to the Appendix, and Sidney Orlov for her valuable editorial assistance.

## Appendix

Here, courtesy of S. Hacıömeroğlu and Y. K. Semertzidis, are parameter definitions and updates [4] of the simulations referenced in [1]. These were the simulations reported in their Storage Ring EDM Collaboration Note (15 June 2011), "SCT with 4th Order Runge-Kutta Simulations" [3]. In parentheses following each figure number, below, is the number of the corresponding figure in [3].

A later report by Hacıömeroğlu and Semertzidis (2014) appeared in NIMA [10]. My Figs. 1 and 5 correspond to Figs. 9 and 5, respectively, of that report.

*Definition of parameters*. The definitions of the plotted quantities are as follows:

$$Y_1 = \frac{q}{mc} \beta_0 E_0 \int \left[ a - \left( \frac{m}{p} \right)^2 \right] dt$$

$$Y_2 = \left[ \vec{\beta} \times \vec{E} \right]_z$$

$$Y_3 = \left[ \vec{\omega}_a \right]_z$$

$$Y_4 = \left[ \int \vec{\omega}_a \, dt \right]_z$$

$\alpha$ = the angle between the spin and momentum vectors as calculated step-by-step during the simulations

$\Delta x$ = radial position

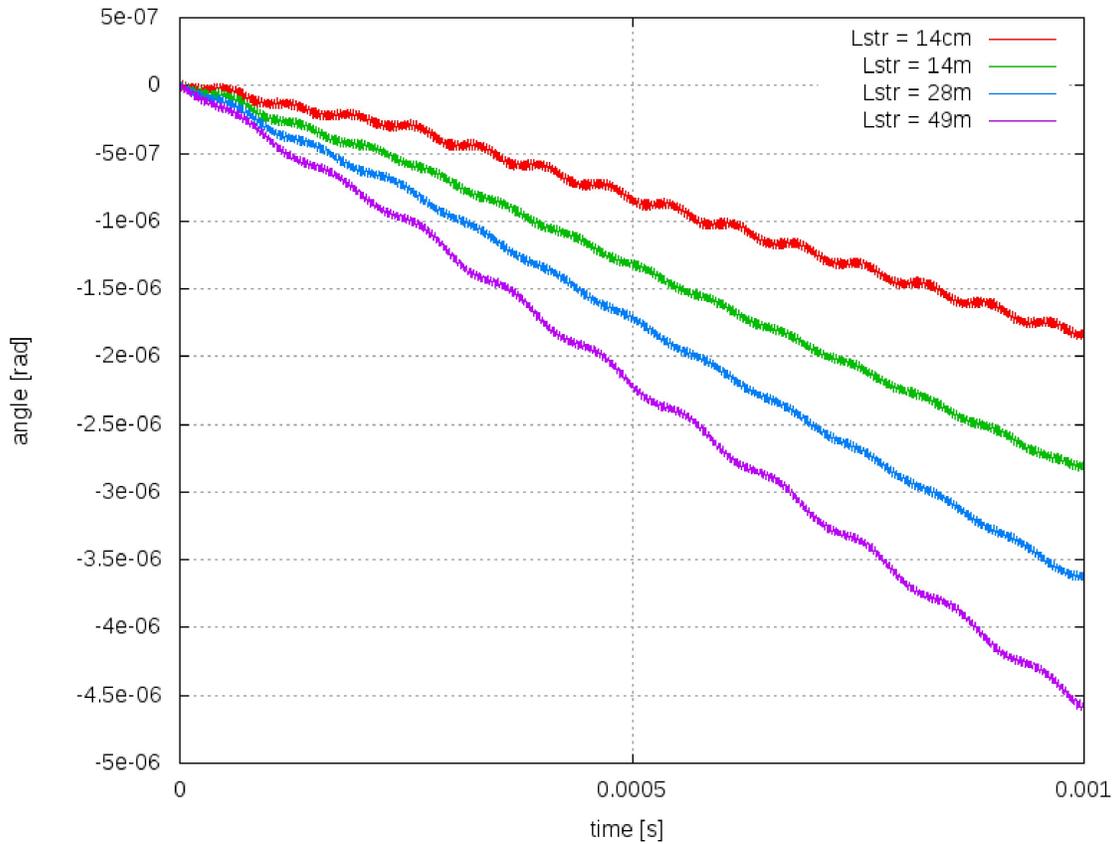

**Fig. 1 (7).** $z_0 = 2\,cm, x_0 = 0, \Delta p/p = 0$. Angle between spin and momentum as a function of time for the $z_0 = 2\,cm$ case. The spin precession rate seems to be 1.6 mrad/s for a 14cm straight section, and 4.5 mrad/s for 49m. Note that we now have 4.5 mrad/s for 49m (instead of the 4 mrad/s of [3]).

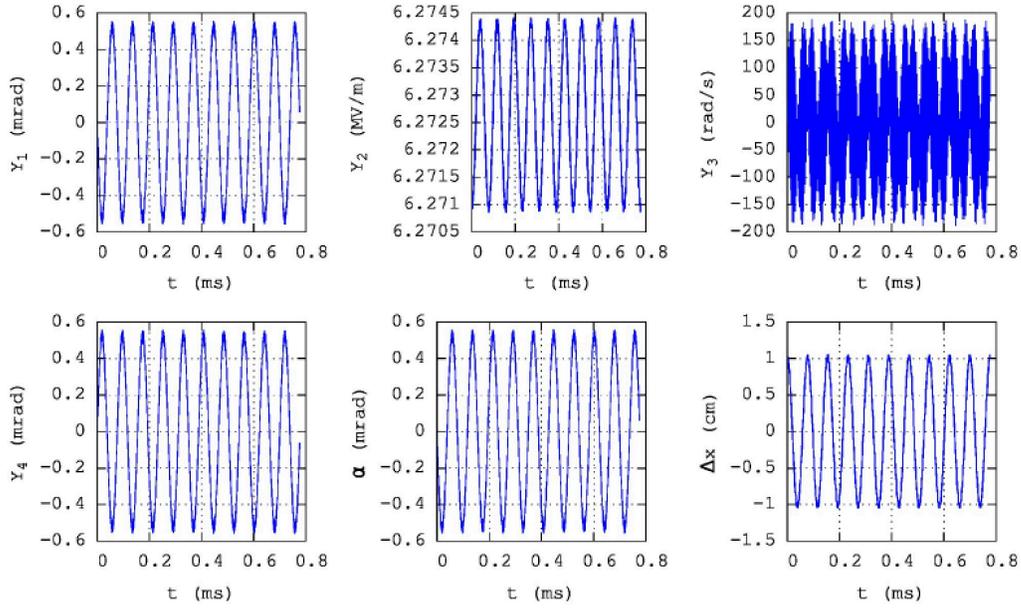

**Fig. 2 (8).** $z_0 = 0, x_0 = 1cm, \Delta p/p = 0$. $x_0$ itself doesn't have a huge effect on SCT.

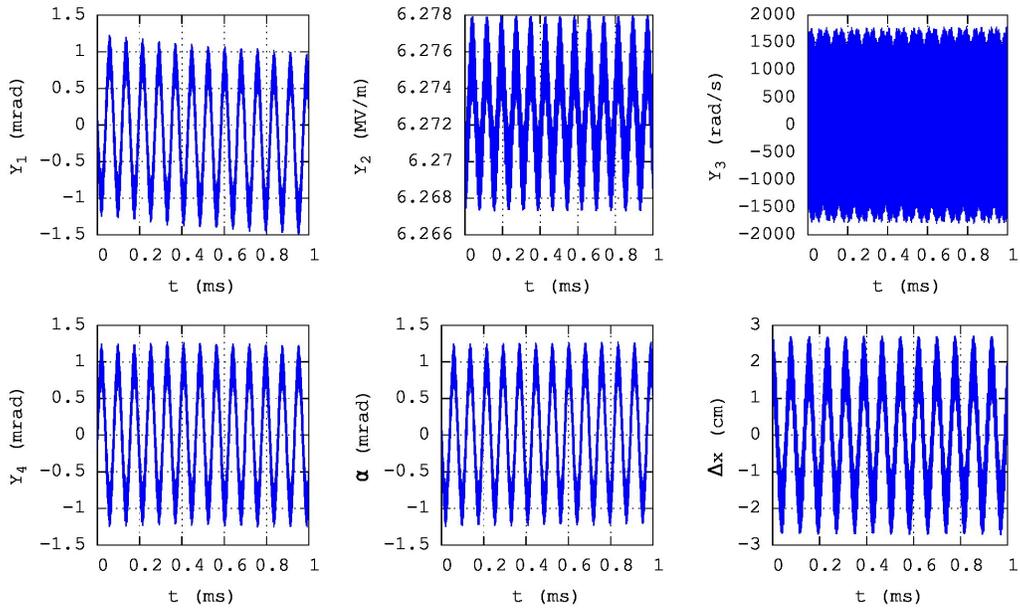

**Fig. 3 (9).** $z_0 = 0, x_0 = 1cm, \Delta p/p = 2 \times 10^{-4}$. Although $Y_1$ tilts by time, $Y_3$ is oscillating constantly because of the $Y_2$ term.

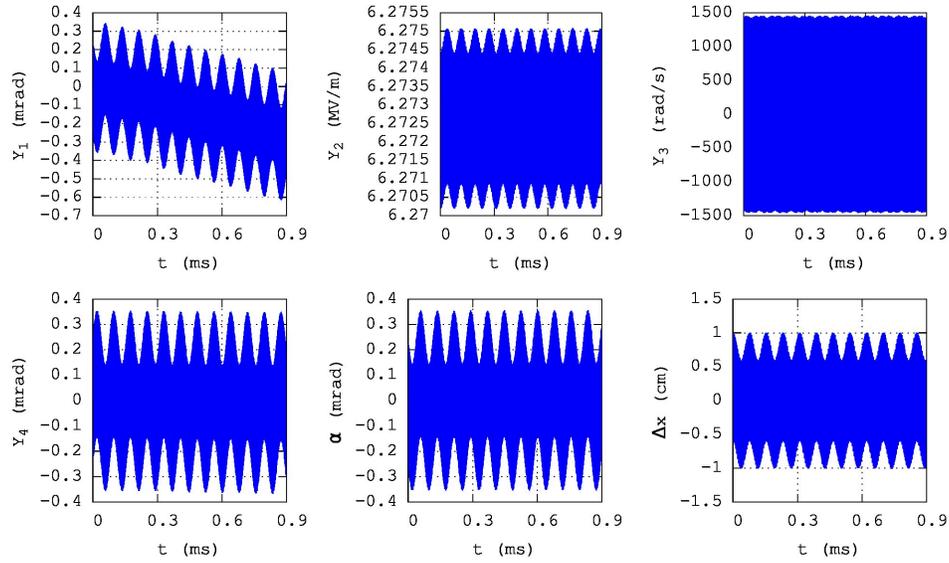

**Fig. 4 (10).** $z_0 = 0, x_0 = 1cm, \Delta p/p = -2 \times 10^{-4}$. Again, $Y_2$ compensates for the tilt in $Y_1$.

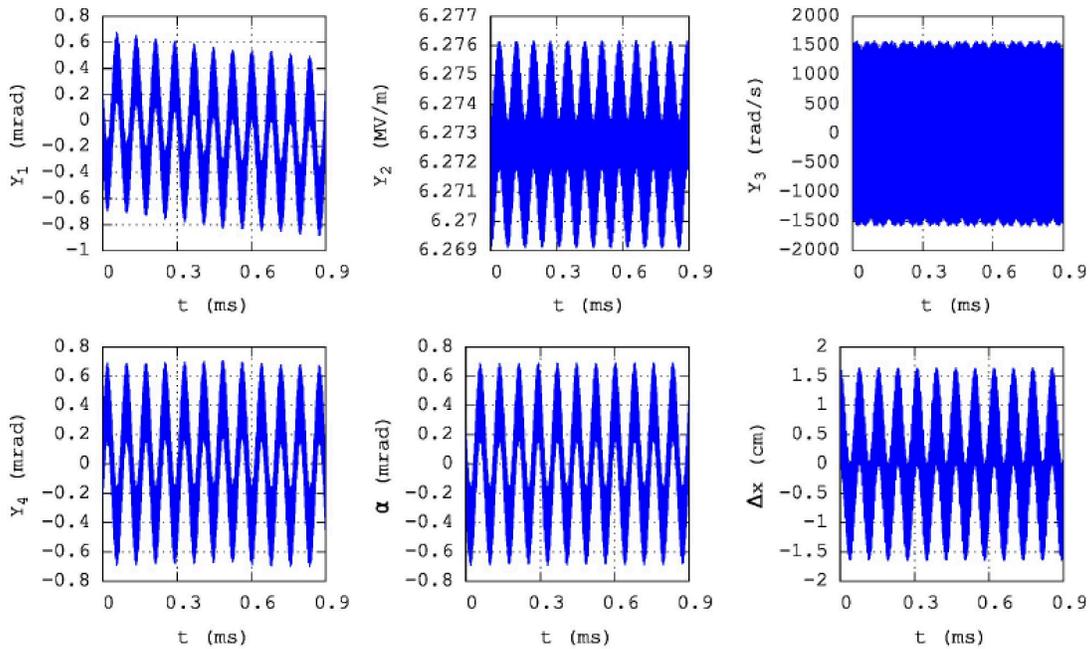

**Fig. 5 (11).** $z_0 = 0, x_0 = 0cm, \Delta p/p = 2 \times 10^{-4}$.